\documentclass[12pt]{article}
\pdfoutput=1
\usepackage{jheppub_X}

\usepackage[utf8]{inputenc}

\usepackage{cleveref}
\crefname{table}{Table}{Tables}
\crefname{equation}{Eq.}{Eqs.}
\crefname{appendix}{App.}{Apps.}
\crefname{section}{Sec.}{Secs.}
\crefname{figure}{Fig.}{Figs.}

\usepackage{setspace}
\usepackage{graphicx}
\usepackage{physics}

\allowdisplaybreaks

\makeatletter
\g@addto@macro\bfseries{\boldmath}
\makeatother

\renewcommand{\a}{\alpha}

\newcommand{\D}{\Delta}

\newcommand{\g}{\gamma}

\renewcommand{\l}{\lambda}

\newcommand{\p}{\pi}

\renewcommand{\r}{\rho}

\newcommand{\f}{\phi}
\newcommand{\F}{\Phi}
\newcommand{\sL}{\mathcal L}
\newcommand{\da}{\dagger}

\title{\huge
A Strongly First-Order Electroweak Phase Transition from Loryons
}

\author[a]{Ian Banta}

\affiliation[a]{Department of Physics, University of California, Santa Barbara, CA 93106, USA}

\emailAdd{banta@ucsb.edu}

\abstract{
We study the effect of BSM particles receiving most of their mass from their coupling to the Higgs boson (``Loryons") on the electroweak phase transition. The existence of BSM Loryons would imply that electroweak symmetry must be non-linearly realized in the effective theory of the Standard Model. Since, by definition, Loryons have a significant coupling to the Higgs, they are expected to have a significant effect on the Higgs effective potential and thereby the electroweak phase transition. We show that the BSM Loryon parameter space viable under current experimental and theoretical constraints overlaps heavily with the parameter space in which a strongly first-order phase transition is predicted. The portion of the experimentally allowed parameter space which gives a strongly first-order phase transition is significantly larger for Loryons as compared to non-Loryons.
}

\begin{document}
\maketitle
\flushbottom
\setcounter{page}{2}

\section{Introduction}
The electroweak phase transition (EWPT), the point in the universe's history when the Higgs boson acquires a vev and the $SU(2)_L\times U(1)_Y$ electroweak symmetry is broken to the $U(1)_\text{em}$ symmetry of electromagnetism, is of great physical interest. It is possible that the matter-antimatter asymmetry observed today is created during the EWPT, a phenomenon termed electroweak baryogenesis (EWBG) \cite{Kuzmin:1985mm,Shaposhnikov:1987tw,Cline:2006ts,Morrissey:2012db}. Doing so requires that the three Sakharov conditions \cite{Sakharov:1967dj} be satisfied at the EWPT. It is also possible for the EWPT to source a stochastic background of gravitational waves (GW), which could be detected by a near-future GW observatory such as LISA \cite{Caprini:2015zlo,Caprini:2019egz}, BBO \cite{Crowder:2005nr}, or DECIGO \cite{Seto:2001qf}. Both of these possibilities require the EWPT to be first-order and sufficiently strong. This is not the case in the Standard Model (SM), as the Higgs mass is too small to give a first-order EWPT \cite{Kajantie:1995kf}. However, by extending the SM with additional particles which couple to the Higgs, the Higgs potential can be modified to give a strongly first-order phase transition, potentially enabling both EWBG\footnote{Sufficient CP-violation is one of the Sakharov conditions which must be satisfied for EWBG, and a fully viable theory of EWBG requires additional CP-violation; this can be added independently through an additional extension of the SM, and in this work we consider only making the EWPT strongly first-order.} and the creation of a stochastic GW background. In addition to the possibility of an observable GW background, such extensions are generally also potentially observable at future colliders through modification of Higgs properties \cite{Katz:2014bha,Ramsey-Musolf:2019lsf}. Much work has explored these possibilities; see, e.g., \cite{Espinosa:1993bs,Barger:2008jx,Ashoorioon:2009nf,Cohen:2012zza,Patel:2012pi,Profumo:2014opa,Curtin:2014jma,Jiang:2015cwa,Kotwal:2016tex,Huang:2016cjm,Vaskonen:2016yiu,Dorsch:2016nrg,Beniwal:2017eik,Kurup:2017dzf,Chiang:2017nmu,Niemi:2018asa,Beniwal:2018hyi,Bell:2019mbn,Kozaczuk:2019pet,Bell:2020gug,Baum:2020vfl,Bell:2020hnr,Cline:2021iff,Schicho:2021gca,Niemi:2021qvp}.\\
\\
In recent work, we examined the phenomenological viability of beyond the Standard Model (BSM) particles which receive most of their mass from the vev of the Higgs boson (``Loryons") \cite{Banta:2021dek}. Such particles are a notable class because the low energy effective theory obtained by integrating out Loryons must be described using HEFT, which linearly realizes only $U(1)_\text{em}$; it cannot be described using SMEFT, which linearly realizes $SU(2)_L\times U(1)_Y$ \cite{Cohen:2020xca,Alonso:2015fsp,Alonso:2016oah,Falkowski:2019tft}. We found a number of examples of Loryons which are consistent with current constraints. By definition, Loryons have a sizable coupling to the Higgs and thus make a significant contribution to the Higgs effective potential. In this work, we leverage the Loryon parametrization and the experimental constraints from \cite{Banta:2021dek} to determine where in the experimentally viable parameter space the EWPT is strongly first-order. We both examine the Loryon parameter space to investigate how prevalent it is to have a strongly first-order EWPT and extend our analysis beyond Loryons to see how relevant the category of Loryons is in picking out extensions which furnish a strongly first-order EWPT.\\
\\
Apart from the focus on Loryons, this paper also improves on previous analyses. As in \cite{Banta:2021dek}, we consider a complete enumeration of viable scalars and vector-like fermions, subject to some assumptions, as opposed to picking out a subset of benchmark models. We have carefully considered direct search bounds on the variety of possible Loryons as well as used the most up-to-date collider constraints and projections, thus providing a more comprehensive and detailed analysis of the constraints. In addition to these, we also impose the partial wave unitarity bounds in \cite{Banta:2021dek} to give an additional constraint. While previous studies (e.g. \cite{Curtin:2014jma}) have included some consideration of unitarity bounds, our treatment is more thorough and leads to a significantly stronger constraint. Exceeding this constraint does not mean a model is absolutely ruled out, but it does mean perturbation theory cannot be relied on to give accurate answers. Finally, much previous work has analyzed the phase transition by focusing on the dynamics at the critical temperature; as discussed in \cite{Kurup:2017dzf,Baum:2020vfl}, it is better and makes a significant difference to calculate at the nucleation temperature, and we do so here.\\
\\
The rest of this paper is organized as follows: In Sec. \ref{sec:catalog}, we discuss the Loryons we consider in this work; this primarily carries over the Loryons found to have viable regions of parameter space in \cite{Banta:2021dek}. In Sec. \ref{sec:calcpt}, we describe our process for determining whether a particular model possesses a strongly first-order EWPT and sources a stochastic GW background. In Sec. \ref{sec:results}, we then show for specific models where in their parameter space they predict a strongly first-order EWPT and detectable GW background. Finally, we present our conclusions in Sec. \ref{sec:conclusion}.

\section{Loryons Considered}
\label{sec:catalog}
In \cite{Banta:2021dek}, we examined the experimental viability of broad classes of Loryons and found that a number of possibilities remain consistent with current constraints. We considered scalars and vector-like fermions; we required that there be no new custodial symmetry violation to the one-loop level, meaning that our Loryons are representations of custodial symmetry $SU(2)_L\times SU(2)_R$; we considered only positive mass-squared terms for new scalars; we imposed a $\mathbb Z_2$ symmetry on the new Loryons, which could be weakly broken to allow decay; and we required that all new charged particles promptly decay. All of these restrictions could be relaxed; we imposed them to simplify the necessary calculations while still capturing the main qualitative points. Relaxing these restrictions generally imposes additional experimental constraints from sources such as custodial symmetry violation, tree-level effects on Higgs couplings, and heavy stable charged particles; with these constraints, only the effects which are an irreducible consequence of coupling to the Higgs and thus necessarily possessed by all Loryons are considered. These restrictions are all maintained in this work. With these restrictions, the potential for a scalar Loryon in an irreducible representation $[L,R]$ of custodial symmetry can include an explicit mass term for the Loryon and a cross-quartic between the Loryon and the Higgs,
\begin{equation}
\sL\supset-\frac{m_\text{ex}^2}{2^\r}\tr(\F^\da\F)-\frac{\l_{h\F}}{2^\r}\tr(\F^\da\F)\frac12\tr(H^\da H).\end{equation}
Here $\r=0(1)$ if the Loryon is complex (real), $m_\text{ex}^2>0$ is an explicit mass parameter, and $\l_{h\F}$ is the cross-quartic coupling. After the Higgs acquires a vev $v$, the physical mass-squared of the Loryon is the sum of contributions from the explicit mass term and the cross-quartic coupling with the Higgs,
\begin{equation}
m_\F^2=m_\text{ex}^2+\frac12\l_{h\F}v^2.
\end{equation}
As discussed in detail in \cite{Banta:2021dek}, integrating out $\F$ to all orders in $H$ gives an effective Lagrangian with terms proportional to one over the Higgs-dependent mass of $\F$,
\begin{equation}
\frac{1}{m_\F^2(H)}=\frac{1}{m_\text{ex}^2+\l_{h\F}\abs{H}^2}.
\end{equation}
A SMEFT description requires expanding about $H=0$ in powers of $\l_{h\F}\abs{H}^2/m_\text{ex}^2$. This is only useful for calculating low energy observables if this series expansion converges at the electroweak breaking vacuum, $\abs{H}=v/\sqrt2$; otherwise, the higher order terms become ever more important rather than ever less important. This requires
\begin{equation}
f_\text{max}\equiv\frac{\l_{h\F}v^2/2}{m_\text{ex}^2}<\frac12.
\end{equation}
i.e. the Loryon must receive less than half its mass from the Higgs. If this condition is not satisfied, implying $f_\text{max}\ge\flatfrac12$, then HEFT (which is an expansion about the electroweak breaking vacuum) is required as the EFT description at low energies\\
\\
In general, one can also write a quartic self-coupling for $\F$ and a second cross-quartic coupling involving contractions through custodial symmetry generators. Different components of $\Phi$ receive different contributions to their mass from this second cross-quartic, inducing some amount of mass splitting. In this case $f_\text{max}$ is the largest value of $f$ over all components; as long as at least one component gets at least half its mass-squared from its coupling to the Higgs, HEFT is required. Appendix A of \cite{Banta:2021dek} includes details of the mass spectrum for different representations. In general, this mass splitting is not qualitatively relevant for this paper and will be set to zero unless otherwise mentioned. Similarly, the quartic self-coupling does not play a significant qualitative role due to our requirement of $m_\text{ex}^2>0$ and the $\mathbb Z_2$ symmetry. We thus set the Loryon self-coupling to the minimum value consistent with perturbative unitarity throughout this paper; including a non-zero self-coupling can shift the mass limits by $\mathcal O(30)$ GeV. The primary model parameters considered are then taken to be $f_\text{max}$ and the physical mass-squared $m_\F^2$.\\
\\
For fermionic Loryons, there is an analogous story. For two vector-like fermions $\Psi_1,\Psi_2$, the Lagrangian can include terms
\begin{equation}
\sL\supset-M_\text{ex1}\tr(\bar\Psi_1\Psi_1)-M_\text{ex2}\tr(\bar\Psi_2\Psi_2)-y_{12}\bar\Psi_1\vdot H\vdot\Psi_2+\text{ h.c.},
\end{equation}
where the custodial representation of the two fermions must be chosen such that the Yukawa term is a singlet. There is once again a Higgs-independent contribution to the mass of the Loryons from the first terms and a Higgs-dependent contribution from the last, with a mass spectrum depending on the precise representations chosen. Here $f_\text{max}$ is defined as the largest fraction of the mass (rather than mass-squared) from the coupling to the Higgs among all components. As detailed in \cite{Banta:2021dek}, $f_\text{max}\ge1/2$ again requires HEFT.\\
\\
A number of experimental constraints were imposed, including Higgs coupling measurements, precision electroweak measurements, and direct searches; see \cite{Banta:2021dek} for details. In addition to these experimental constraints, imposing perturbative unitarity places an upper bound on the mass of new Loryons.\\
\\
For scalar Loryons, a neutral singlet, charged singlet, hypercharge 0 electroweak triplet, hypercharge 1/2 electroweak doublet, and charge 2/3 and -1/3 color triplet are experimentally viable for broad swathes of their parameter space. A few additional scalar Loryons, such as a hypercharge 1 electroweak triplet and hypercharge 1/2 electroweak quadruplet, have small regions of viability. Additional possibilities, including fermionic Loryons and scalar Loryons in larger representations or with larger charges, are experimentally ruled out when added on their own but viable when added in combination with other BSM particles. There are a large number of possibilities which generally have delicate cancellations to evade the experimental constaints; see \cite{Banta:2021dek} for more detailed discussion. Due to the number of possibilities, we do not perform a complete analysis of the parameter space for these cases, but do discuss the general features and compute with particular models to check that these hold.

\section{Calculating the EWPT Characteristics}
\label{sec:calcpt}
The characteristics of the EWPT can be calculated from the effective potential for the Higgs boson. Doing so with a perturbative calculation in finite-temperature field theory has problems such as gauge-dependence, infrared divergences, and others; see, e.g., \cite{Linde:1980ts,Gross:1980br,Patel:2011th,Garny:2012cg,Lofgren:2021ogg,Hirvonen:2021zej}. A non-perturbative calculation using e.g. lattice computations is necessary for a fully accurate result; here, we do not deal with these issues, using standard perturbative techiniques to estimate the properties of the EWPT (see, e.g., \cite{Morrissey:2012db,Kozaczuk:2019pet,Baum:2020vfl,Cline:2021iff}). We expect, given the approximations inherent in doing a perturbative calculation, that the borders of the region we determine to have an appropriate EWPT will be fuzzy; for any particular model, a more precise calculation resolving the issues described above could be carried out to give a more definitive answer. However, our intention is only to characterize the broad regions of parameter space and show the degree to which they overlap with the viable parameter space from the previous study, and for this the perturbative calculation presented here is sufficient.\\
\\
This section is split into three subsections. In Sec.\ \ref{subsec:effpot}, we describe the procedure we use to compute the effective potential for the Higgs boson. In Sec.\ \ref{subsec:firstorder}, we discuss how we use this effective potential to determine whether there is a strongly first-order EWPT. In Sec.\ \ref{subsec:gwbg}, we discuss our criteria for whether an EWPT will source a detectable stochastic background of gravitational waves.

\subsection{The Effective Potential}
\label{subsec:effpot}
Our construction of the effective potential follows standard methods; see, e.g., \cite{Quiros:1999jp}. At zero temperature and to one-loop order, the effective potential can be written
\begin{equation}
V_{\text{eff},T=0}(h,\f)=V_0(h,\f)+\sum_in_iV_\text{CW,b}(m_i^2(h,\f))+\sum_in_iV_\text{CW,f}(m_i^2(h,\f)).
\end{equation}
Here $h$ is the SM Higgs; $\f$ are the BSM Loryons; $V_0$ is the tree-level potential; $V_\text{CW,b/f}$ is the one-loop Coleman-Weinberg correction for bosons/fermions; $n_i$ is the number of degrees of freedom of the particle; and the sum runs over all particles but in practice receives significant contributions only from those with significant coupling to $h,\f$. The one-loop corrections take the form
\begin{equation}
V_\text{CW,b/f}(m_i^2(h,\f))=\pm\frac{1}{64\p^2}m_i^2(h,\f)\left(m_i^2(h,\f)\log(\frac{m_i^2(h,\f)}{m_i^2(v_h,v_\f)})+2m_i^2(v_h,v_\f)\right)
\end{equation}
where the $\pm$ is for bosons/fermions and $(v_h,v_\f)$ are the tree-level vacuum expectation values of $h,\f$; this form has fixed counterterms such that the tree-level relations between the Lagrangian parameters and the Higgs mass and vev are preserved. At finite temperature, there are two corrections. The first is the addition of a one-loop thermal potential to the Coleman-Weinberg term,
\begin{equation}
\D V_{\text{eff},T>0}(h,\f,T)=\sum_in_iV_\text{T,b}(m_i^2(h,\f),T)+\sum_in_iV_\text{T,f}(m_i^2(h,\f),T),
\end{equation}
where
\begin{gather}
V_\text{T,b/f}(m_i^2(h,\f),T)=\frac{T^4}{2\p^2}J_\text{b/f}\left(\frac{m_i^2(h,\f)}{T^2}\right),\\
J_\text{b/f}(y^2)=\pm\int_0^\infty\dd x\,x^2\log(1\mp\exp(-\sqrt{x^2+y^2})).
\end{gather}
The second correction is due to the resummation of hard thermal loops, which are nominally of higher-loop order but give a significant numerical contribution due to the presence of the additional dimensionless ratio $T/m$, where $m$ is a mass scale \cite{Parwani:1991gq,Comelli:1996vm}. This can be handled by making the substitution
\begin{equation}
m_i^2(h,\f)\to m_i^2(h,\f)+\Pi_i(T),
\end{equation}
where $\Pi_i$ is found by differentiating $V_\text{T}$; typically only the non-zero contribution of leading-order in $T$ is kept, for which
\begin{equation}
J_\text{b,lo}\left(\frac{m^2}{T^2}\right)=2J_\text{f,lo}\left(\frac{m^2}{T^2}\right)\approx\frac{\pi^2}{12}\frac{m^2}{T^2},
\end{equation}
and thus all the $\Pi_i$ are directly proportional to $T^2$.\\
\\
The calculation described above is not the state of the art even for a perturbative calculation; it can be improved through methods such as dimensional reduction \cite{Croon:2020cgk}, higher loop order \cite{Cohen:2011ap,Niemi:2021qvp}, and renormalization group improvement \cite{Ramsey-Musolf:2019lsf,Gould:2021oba}. As our intention is only to characterize broad regions of parameter space, we do not improve our calculation in any of these ways. For a particular model of interest, a more precise calculation using these methods would be desirable to obtain a more robust answer.\\
\\
Due to the assumptions of $\mathbb Z_2$ symmetry and positive quadratic terms for the new Loryons, we assume that only the Higgs field acquires a non-zero vacuum expectation value (and have checked models throughout the parameter space to confirm that this assumption holds). This means that we can drop all $\f$ dependence in the above formulae so that the effective potential is a function of just $h,T$.

\subsection{Conditions for a Strongly First-Order EWPT}
\label{subsec:firstorder}
Given an effective potential, we now need to determine whether it gives a strongly first-order transition. As noted earlier, this is a necessary (but not sufficient) condition for electroweak baryogenesis. A first-order transition can occur between two local minima, the symmetry-preserving minimum at $h=0$ and the symmetry-breaking one at $h>0$. At the critical temperature $T_c$, these two minima are degenerate,
\begin{equation}
V_\text{eff}(0,T_c)=V_\text{eff}(h_c,T_c),
\end{equation}
where $h_c>0$ minimizes $V_\text{eff}(h,T_c)$. Below this temperature, the symmetry-breaking minimum becomes the global minimum, and bubbles of the symmetry-breaking minimum can nucleate by tunneling through the barrier between the two minima. The nucleation temperature $T_n$ is defined as the temperature at which the probability of a bubble nucleating within one Hubble volume per Hubble time is $\mathcal O(1)$; this corresponds to $S_3/T\approx140$, where $S_3$ is the three-dimensional Euclidean action of the critical bubble \cite{Quiros:1999jp,Moore:1995si,Wainwright:2011kj}. For the EWPT to be strongly first-order and potentially suitable for electroweak baryogenesis, the Higgs vev in the broken phase must be large enough to suppress sphalerons. We make the standard approximation of using $v_n/T_n$ to measure the strength of the phase transition, with $v_n/T_n\gtrsim1$ the condition for strong enough to be suitable for electroweak baryogenesis \cite{Quiros:1999jp}. We then make a rough estimate of the precision of our computation by varying the thresholds for $S_3/T$ and $v_n/T_n$. For $S_3/T$ we vary by a factor of 1.4, computing with a threshold of $S_3/T=100,200$; for $v_n/T_n$ we vary by a factor of 1.6, computing with $v_n/T_n>.6,1.6$.\\
\\
In the Standard Model, the EWPT is a crossover, not a first-order transition \cite{Kajantie:1995kf}; the minimum moves smoothly away from $h=0$ rather than having two degenerate local minima and tunneling between them. Adding new particles coupled to the Higgs can lower the symmetry-breaking minimum relative to the symmetry-preserving minimum and enhance the barrier between the two minima, leading to a first-order transition. Requiring a strongly first-order EWPT thus requires some minimum contribution from BSM physics; in Sec.\ \ref{sec:results}, we will see this play out as requiring sufficiently large mass and $f_\text{max}$. There is also a maximum workable contribution; adding too many new scalars too strongly coupled to the Higgs can stabilize the Higgs effective potential at the symmetry-preserving minimum in violation of the observed zero-temperature symmetry breaking. We impose as a condition $T_n>10$ GeV; the location of the upper bound in the parameter space is only weakly sensitive to the precise condition imposed as long as it is below $\sim50$ GeV.

\subsection{Gravitational Wave Background}
\label{subsec:gwbg}
We are also interested in when the EWPT sources a GW background which could be detected by a near-future GW observatory such as LISA, BBO, or DECIGO. GW are primarily sourced from the EWPT through collision of bubble walls, sound waves, and turbulence during and after collisions of bubbles of broken phase \cite{Caprini:2015zlo}. It is possible to calculate spectra for the GW and compare these to sensitivity curves; see, e.g., \cite{Wainwright:2011qy,Caprini:2015zlo,Breitbach:2018ddu,Gould:2019qek,Alanne:2019bsm,Caprini:2019egz,Wang:2020jrd}. An approximate answer can be obtained based on only two properties of the phase transition: the ratio of the vacuum energy density released in the transition to the radiation energy density,
\begin{equation}
\a=\flatfrac{\left(\D V_\text{eff}-\frac{T_n}{4}\D\dv{V_\text{eff}}{T}\right)}{\frac{g_*\pi^2T_n^4}{30}},
\end{equation}
and the inverse duration of the phase transition over the Hubble parameter \cite{Caprini:2015zlo,Caprini:2019egz},
\begin{equation}
\beta/H_*=\left.\dv{S_3}{T}\right|_{T_n}-\frac{S_3}{T_n}.
\end{equation}
Physically, both a larger energy release in transitioning out of the metastable minimum (large $\a$) and a longer duration (small $\beta$) source a stronger GW signal. Both large $\a$ and small $\beta$ tend to apply for stronger transitions; thus, the regions of the parameter space with the strongest transitions are expected to produce a detectable GW background. Based on \cite{Alanne:2019bsm,Caprini:2019egz}, we use
\begin{equation}
\log(\beta/H_*)\lesssim1.2\log\a+8.8
\end{equation}
as the condition for the phase transition to source a GW background detectable by LISA. The cutoff for where in the model parameter space we expect a detectable GW background is only weakly dependent on the condition chosen; adjusting the required value of $\beta/H_*$ by an order of magnitude only shifts the bound on the Loryon mass by $\mathcal O(3)$ GeV. For BBO and DECIGO, we estimate from \cite{Breitbach:2018ddu} that the constant 8.8 should be increased to around 9.2. Again we emphasize that this is an approximate estimate, but the end result as displayed in the Loryon parameter space is only weakly dependent on the condition chosen. For a more precise estimate for a particular model, a more detailed calculation would be needed; however, as stated before, we are interested only in characterizing broad regions of parameter space, for which this estimate is sufficient.

\section{Results}
\label{sec:results}

\begin{figure}
\centering
\includegraphics[width=\textwidth]{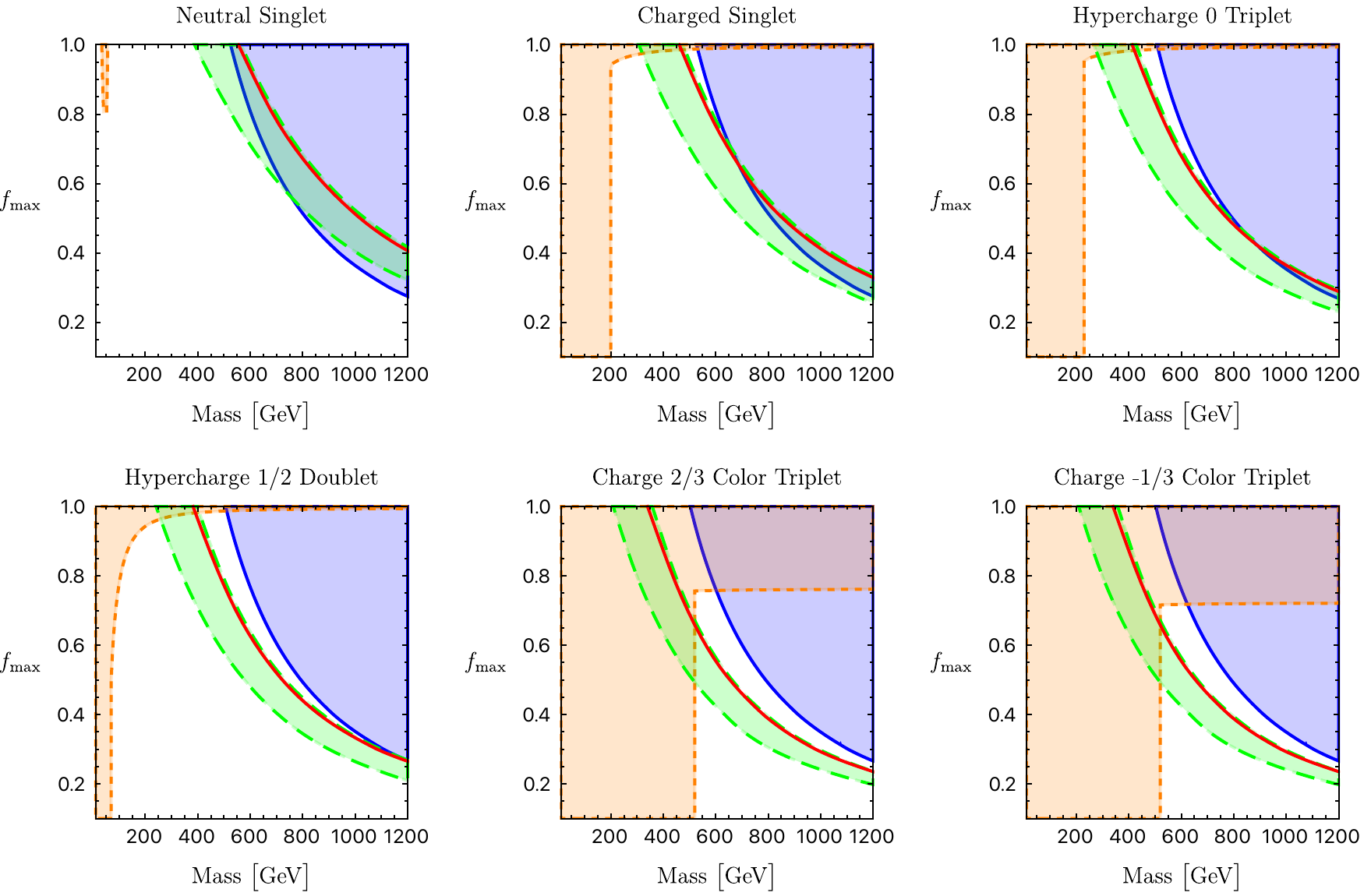}\hspace{55pt}
\caption{The allowed parameter space for which there is a strongly first-order electroweak phase transition. The mass quoted is the physical mass of the Loryon. $f_\text{max}$ is the fraction of its mass-squared the particle gets from its coupling to the Higgs; $f_\text{max}>1/2$ is the condition for a Loryon. The orange region (dotted boundary) is ruled out by experimental constraints; the blue region (solid boundary) is ruled out by perturbative unitarity. The green region (dashed boundary) is where a strongly first-order phase transition is predicted. The solid red line through this region is the lower bound for production of a stochastic gravitational wave background detectable by LISA, which is only marginally below the upper bound for an appropriate phase transition.}
\label{fig:parspace}
\end{figure}

Results for a selection of viable Loryons are shown in Fig. \ref{fig:parspace}. We see that the region of parameter space in which there is a strongly first-order EWPT overlaps heavily with the region of parameter space consistent with current experimental and theoretical constraints. There is a much smaller region for which the EWPT sources a GW background detectable by near-future GW observatories; however, this also lies within the experimentally viable parameter space for all models except the neutral singlet Loryon.\\
\\
For purposes of comparison, the plots show the parameter space through smaller values of $f_\text{max}$, going below the cutoff for Loryons of $f_\text{max}=1/2$. The unitarity bound increases sharply outside the Loryons regime, as does the band with a viable phase transition. It is thus not required to add Loryons in order to achieve a strongly first-order EWPT; however, the viable region overlaps with a larger portion of the experimentally viable parameter space for Loryons as compared to non-Loryons, with (typically) $\sim$30-50\% of the experimentally viable region producing a suitable phase transition for Loryons and $\sim$10\% for non-Loryons. These estimates vary significantly with $f_\text{max}$ and with the representation chosen; we refer the reader to Fig. \ref{fig:parspace} for more detailed information. The larger viable region for Loryons is not an artifact of the chosen parametrization of $(m,f_\text{max})$; if viewed in terms of the cross-quartic between the Higgs and the new field rather than the mass, smaller values of $f_\text{max}$ require a cross-quartic larger by a factor of $\mathcal O(3)$, while the range of cross-quartics giving a viable EWPT remains approximately the same size.\\
\\
The plots are shown for our standard values of $S_3/T=140$, $v_n/T_n>1$. Varying the value of $S_3/T$ shifts the bounds by $\mathcal O(3)$ GeV, a negligible effect. Varying the value of $v_n/T_n$ has a stronger effect; a factor of 1.6 variation shifts the lower bound of the viable region by $\mathcal O(50)$ GeV. Increasing the Loryon self-coupling increases both the upper and lower bound by $\mathcal O(30)$ GeV. The curve for a detectable GW background is drawn using our estimate for LISA; for BBO or DECIGO, the curve would shift to the left by $\mathcal O(5)$ GeV.\\
\\
As shown in Fig. \ref{fig:parspace}, larger representations have the region for a strongly first-order phase transition shifted to lower masses. This is due to the greater number of degrees of freedom in the larger representations. The viable region depends principally on the number of degrees of freedom added; the exact electroweak representation has only a small effect on the region for which there is a viable phase transition. One can thus read the first four panels of \ref{fig:parspace} as showing the approximate viable region for adding 1, 2, 3, and 4 real scalar degrees of freedom; adding 5, 6, etc. would continue to shift the viable region to lower masses. The principal experimental constraint, due to Higgs decay to two photons, is that there may be at most one charged particle, so additional Loryons beyond the representations given would need to be neutral singlets. Of course, there are a larger number of parameters when adding Loryons in multiple representations, as the different representations can have different masses and values of $f_\text{max}$; interpreting the plots as showing the viable region for an appropriate number of new fields in different representations is projecting the allowed region into a slice wherein all new particles have the same mass and coupling to the Higgs.\\
\\
In \cite{Banta:2021dek}, we also discussed the possibility of fermionic Loryons and scalar Loryons in larger representations being viable by adding enough charged BSM particles; this flips the sign of the Higgs coupling to two photons but matches the magnitude of the SM \cite{Bizot:2015zaa}. There are a number of possibilities for doing so and we have not exhaustively characterized where in their parameter spaces one predicts a strongly first-order EWPT; however, a few general comments are in order. Adding too many scalar degrees of freedom renders the symmetric minimum too stable. How many is too many depends on the mass and fraction of that mass from the Higgs; as an example, only up to 6 electroweak quadruplet scalars getting half their mass-squared from the Higgs with mass 400 GeV can be added in isolation, while 8 would be required to flip the $h\gamma\gamma$ coupling. It is thus not possible to add only scalars while flipping the $h\gamma\gamma$ coupling. The one-loop correction to the effective potential from fermions is of opposite sign to that from scalars, and so it is possible to add a mix of scalar and fermionic Loryons and get the desired behavior for the EWPT. While the parameter space is too extensive to perform a broad scan, we have checked particular examples in detail to confirm that we can in principle get a strongly first-order phase transition. For a model with an appropriate mix of scalars and fermions, there is a non-trivial mass range at a mass of several hundred GeV for which we predict a strongly first-order electroweak phase transition.\\
\\
\begin{figure}
\centering
\includegraphics[width=\textwidth]{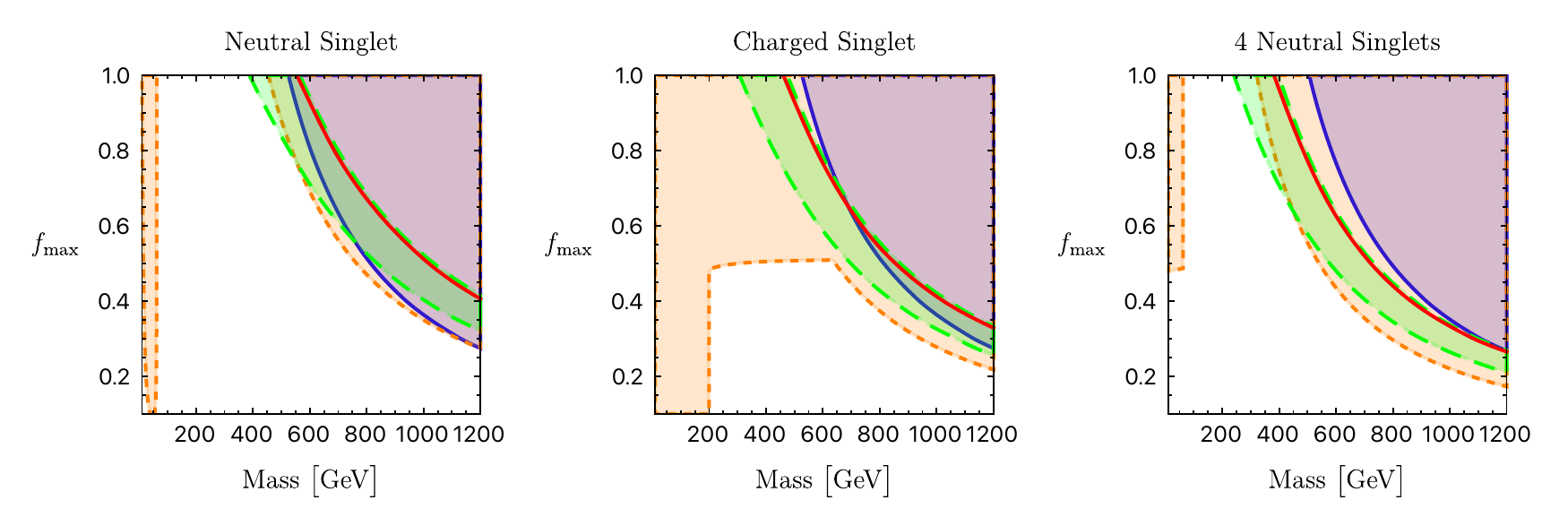}\hspace{55pt}
\caption{The allowed parameter space based on the projections for near-future colliders. We use the one-sigma projections for the FCC-ee at 240 GeV from \cite{deBlas:2019rxi}, doubled to give an estimate of the two-sigma bound; other colliders, such as ILC and CLIC, have comparable projections \cite{deBlas:2019rxi}. Labels are as Fig. \ref{fig:parspace}. The two principal constraints are $h\g\g$, which comes down from the top in the middle panel, for which the projected two-sigma constraint is $2.6\%$; and the Higgs cubic self-coupling, which comes in from the left in the figure, for which the projected two-sigma constraint is $38\%$.}
\label{fig:FCCparspace}
\end{figure}\\
\\
In Fig. \ref{fig:FCCparspace}, we show the viable parameter space with projected two-sigma constraints from the FCC-ee; other near-future colliders, such as ILC and CLIC, have similar projections \cite{deBlas:2019rxi}. The measurement of the $h\g\g$ interaction is projected to be precise enough to rule out charged and colored Loryons, leaving some number of neutral singlets as the only viable possibility. The Higgs cubic self-coupling then places a stringent constraint; we predict that achieving a strongly first-order EWPT while evading the constraint from the Higgs cubic would require singlets getting at least $\sim.7$ of their mass-squared from the Higgs and occupying a narrow mass range. However, recall that our calculations of the viable region are approximate, with an estimated error of $\mathcal O(50)$ GeV. This is enough for the boundary of the viable region to move substantially relative to the projected Higgs cubic constraint, and so a more precise calculation of the phase transition would be required for definitive conclusions. Nevertheless, it is likely that a model of Loryons giving a strongly first-order EWPT would furnish a detectable signal at a near-future collider such as the FCC-ee. Additionally, a model which produces a stochastic GW background would be expected to be observable at a near-future collider.

\section{Conclusions}
\label{sec:conclusion}
In this work we have examined particles getting most of their mass from their coupling to the Higgs field (``Loryons"), subject to some modest assumptions, and their potential effect on the electroweak phase transition. As Loryons by definition have a significant coupling to the Higgs, it is reasonable to expect that they have a significant effect on the Higgs effective potential and can enable a strongly first-order EWPT. For BSM Loryons which are viable when added by themselves, we performed scans over the parameter space and calculated the properties of the EWPT. We find that the region of parameter space in which the Loryons are experimentally viable overlaps heavily with the parameter space for which a strongly first-order phase transition is predicted.\\
\\
Loryons cut off at $f_\text{max}=1/2$, as this is the point at which a BSM particle requires HEFT for an effective field theory description. The viable band has generally not yet run into the unitarity bound by $f_\text{max}=1/2$, and so it is not required to add Loryons to achieve a strongly first-order EWPT.  However, beyond the Loryon regime the viable band turns sharply to higher masses, so the portion of the experimentally allowed parameter space which gives a strongly first-order phase transition is significantly larger for Loryons as compared to non-Loryons. This is not an artifact of the chosen parametrization of $(m,f_\text{max})$; if viewed in terms of the cross-quartic between the Higgs and the new field rather than the mass, smaller values of $f_\text{max}$ require a cross-quartic larger by factor of $\mathcal O(3)$, while the range of cross-quartics giving a viable EWPT remains approximately the same size.\\
\\
As discussed in \cite{Banta:2021dek}, there are Loryons which are not viable on their own but which can satisfy the experimental constraints when added with enough appropriate other Loryons. For these cases, not as much of the parameter space gives an appropriate EWPT since adding too many new scalar Loryons stabilizes the effective potential at the origin. However, when both scalar and fermionic Loryons are added there are still broad ranges of the parameter space which would predict a strongly first-order EWPT.\\
\\
The parameter space for which the EWPT produces a stochastic gravitational wave background detectable by near-future observatories is much smaller. However, it once again overlaps with the experimentally viable parameter space. Of particular note, however, is that it does not overlap with the viable parameter space for the neutral singlet; we conclude that the bound from imposing partial wave unitarity on the $S$-matrix is sufficient to rule out a detectable stochastic GW background from a (one-step) first-order EWPT driven by a neutral singlet. The $\mathbb Z_2$-symmetric neutral singlet is popularly referred to as a ``nightmare scenario" due to its paucity of collider signals, and this would mean that a viable model obeying our assumptions could not be detected by near-future GW observatories. However, a few points of caution are in order. First, as stated before, we have made only an approximate calculation of the phase transition and its characteristics; nevertheless, our result is consistent with more detailed calculations in the literature such as \cite{Beniwal:2017eik,Caprini:2019egz}. Second, we have only considered the possibility of a one-step phase transition, with no vev for the singlet; the partial wave unitarity bound does not rule out the region of parameter space with a two-step transition and a detectable GW background. Third, only a model with a single neutral singlet is ruled out in this way; the partial wave unitarity is only slightly affected by adding more singlets, while the region of parameter space with a detectable GW background shifts down.\\
\\
Due to their coupling to the Higgs, new Loryons would have a significant effect on Higgs properties and could be detected via this avenue at future colliders. Improveed measurement of $h\g\g$ could detect or rule out charged Loryons. Measurements of the Higgs cubic self-coupling are projected to overlap heavily with the region predicted to have a strongly first-order phase transition, with a measurement of order tens of percent probing much of this region and potentially leading to a discovery. For a precision such as that projected for the FCC-ee, ILC, or CLIC, a more detailed calculation of the phase transition would be required to give precise results, but we expect only Loryons getting at least $\sim.7$ of their mass-squared from the Higgs in a narrow mass range could provide a strongly first-order EWPT while evading detection via modifications to the Higgs cubic. In particular, this means that the viable region for non-Loryons could be entirely ruled out.\\
\\
We imposed a $\mathbb Z_2$ symmetry on the new Loryons and required that they have positive mass-squared terms. It is possible to relax these assumptions; doing so would be particularly notable for EWBG as it would allow the possibility of a two-step transition where one of the new fields acquires a vev. Relaxing these assumptions leads to tree-level signatures and thus additional experimental constraints on the new Loryons, but it would be interesting to examine this broader parameter space for the existence of a viable EWPT.\\
\\
The calculations herein have only characterized broad regions of parameter space. For models of particular interest, one could make a more precise calculation of the effective potential, the phase transition, and its properties. In addition, we have only examined the question of whether the phase transition is strongly first-order. While this is a necessary condition for electroweak baryogenesis, it is not the only one. Future work could develop a model with additional CP-violating ingredients added to one or more BSM Loryons to give a full theory of electroweak baryogenesis. Similarly, more precise calculations could be done for specific models of interest to confirm their possible detectability via gravitational wave backgrounds.\\
\\
BSM particles which receive most of their mass from couplings to the Higgs boson remain experimentally viable. Thanks to this coupling, they not only have novel, detectable collider phenomenology but also have a strong effect on the physics of the electroweak phase transition, may contribute to baryogenesis, and could potentially be detected through gravitational wave observation.

\acknowledgments

Thanks to Timothy Cohen, Xiaochuan Lu, and Dave Sutherland for useful conversations, and to Nathaniel Craig for useful conversations and assisting in the preparation of the manuscript. The author is supported by the U.S.~Department of Energy under the grant DE-SC0011702.

\addcontentsline{toc}{section}{\protect\numberline{}References}%
\bibliographystyle{JHEP}
\bibliography{Loryon_EWPT_JHEP_1}

\end{document}